\documentclass[twocolumn,showpacs,preprintnumbers,amsmath,amssymb,superscriptaddress]{revtex4}
\usepackage{graphicx}
\usepackage{dcolumn}
\usepackage{bm}
\usepackage{longtable}

\newcommand{\bea}{\begin{eqnarray}}
\newcommand{\eea}{\end{eqnarray}}

\usepackage{amsfonts}
\usepackage{amssymb}
\usepackage{amsmath}
\usepackage{amsthm}
\usepackage{graphicx}

\newcommand{\hi}{\mathcal{H}}

\newcommand{\be}{\begin{equation}}
\newcommand{\ee}{\end{equation}}

\def\b{\beta}

\def\d{\delta}

\def\Be'{\beta_\mu^{'}}

\def\<{\bigl\langle}
\def\>{\bigr\rangle}

\def\ll{\langle}
\def\rr{\rangle}

\begin{document}


\title{Anergy in self-directed B lymphocytes from a statistical mechanics perspective.}

\author{Elena Agliari}
\affiliation{Dipartimento di Fisica, Universit\`{a} degli Studi di Parma, viale G. Usberti 7, 43100 Parma, Italy}
\affiliation{INFN, Gruppo Collegato di Parma, viale G. Usberti 7, 43100 Parma, Italy}
\author{Adriano Barra}
\affiliation{Dipartimento di Fisica, Sapienza Universit\`{a} di Roma, Piazzale Aldo Moro 2, 00185, Roma, Italy}
\author{Gino Del Ferraro}
\affiliation{Department of Computational Biology, Royal Institute of Technology, SE-100 44, Stockholm, Sweden}
\author{Francesco Guerra}
\affiliation{Dipartimento di Fisica, Sapienza Universit\`{a} di Roma, Piazzale Aldo Moro 2, 00185, Roma, Italy}
\affiliation{INFN, Gruppo di Roma, Piazzale Aldo Moro 2, 00185, Roma, Italy}
\author{Daniele Tantari}
\affiliation{Dipartimento di Matematica, Sapienza Universit\`{a} di Roma, Piazzale Aldo Moro 2, 00185, Roma, Italy}

\date{\today}

\begin{abstract}
The ability of the adaptive immune system to discriminate between self and non-self mainly stems from the ontogenic
clonal-deletion of lymphocytes expressing strong binding affinity with self-peptides. However, some self-directed lymphocytes may evade selection and still be harmless due to a mechanism called clonal anergy.

As for B lymphocytes, two major explanations for anergy developed over three decades: according to "Varela theory", it stems from a proper orchestration of the whole B-repertoire, in such a way that self-reactive clones, due to intensive interactions and feed-back from other clones, display more inertia to mount a response. On the other hand, according to the `two-signal model", which has prevailed nowadays, self-reacting cells are not stimulated by helper lymphocytes and the absence of such signaling yields anergy.

The first result we present, achieved through disordered statistical mechanics, shows that helper cells do not prompt the activation and proliferation of a certain sub-group of B cells, which turn out to be just those broadly interacting, hence it merges the two approaches as a whole (strictly speaking Varela theory is then included into the two-signal model, not vice-versa).

As a second result, we outline a minimal topological architecture for the B-world, where highly connected clones are self-directed as a natural consequence of an ontogenetic learning; this provides a mathematical framework to Varela perspective.\\
As a consequence of these two achievements, clonal deletion and clonal anergy can be seen as two inter-playing aspects of the same phenomenon too.

\end{abstract}



 \maketitle

\section{Introduction}

The adaptive response of the immune system is performed through the coordination of a huge ensemble of cells (e.g. B cells, helper and regulatory cells, etc.), each with specific features, that interact both directly and via exchanges of chemical messengers as cytokines and immunoglobulins (antibodies) \cite{janaway}.
In particular, a key role is played by B cells, which are lymphocytes characterized by membrane-bound immunoglobulin (BCR) working as receptors able to specifically bind an antigen; upon activation, B cells produce specific soluble immunoglobulin.
B cells are divided into clones: cells belonging to the same clone share the same specificity, that is, they express the same BCR and produce the same antibodies (hyper-somatic mutations apart \cite{janaway}). When an antigen enters the host body, some of its fragments are presented to B cells, then, the clones with the best-matching receptor, after the authorization of helpers through cytokines, undergo clonal expansion and release a huge amount of antibodies in order to kill pathogens and restore order.

This picture, developed by Burnet \cite{burnet} in the $50$'s  and verified across the decades, constitutes the ``clonal selection theory'' and, when focusing on B-cells only, can be looked at as a one-body theory \cite{BA1}: The growth (drop) of the antigen concentration elicits (inhibits) the specific clones.

One step forward, in the $70$'s, Jerne suggested that, beyond antigenic stimulation, each antibody must also be detected and acted upon by other antibodies; as a result, the secretion of an atypically large concentration of antibodies  by an active B clone (e.g. elicited due to an antigen attack) may even prompt the activation of other B clones that best match those antibodies \cite{jerne}. This mechanism, experimentally well established (see e.g. \cite{cazenave, lanzetta}), underlies a two-body theory and (possibly) gives rise to an effective network of clones interacting via antibodies, also known as ``idiotypic network''.


The B repertoire is enormous ($\sim 10^9$ in humans) and continuously updated due to the random gene-reshuffling occurring during B-cell ontogenesis in the bone marrow \cite{janaway}. The latter process ensures the diversity of the repertoire and therefore the ability of the immune system to recognize many different antigens, but, on the other hand, it also inevitably produces cells able to detect and attack self-proteins and this possibly constitutes a serious danger.
In order to avoid the release of such auto-reactive cells, safety mechanisms are at work during the ontogenesis, yet, some of them succeed in escaping through ``receptor editing'' (self-reactive cells substitute one of their receptors on their immunoglobulin surface) \cite{kitamura} or ''clonal anergy'' (self-reactive cells that have not been eliminated or edited in the bone marrow become unresponsive, showing reduced expression level of BCR) \cite{goodnow1,goodnow2}.

In the last decades, two main strands have been proposed to explain clonal anergy, both supported by experimental evidence: The former, introduced by Varela \cite{varela2,varela3,varela1}, allows for B cells only, while the latter, referred to as the two-signal model \cite{goodnow1,goodnow2,goodnow3}, allows for both B and helper T cells.

According to Varela's theory, each clone $\mu$ corresponds to a node in the idiotypic network, with a (weighted) coordination number $W_{\mu}$ (i.e. the sum of the binding strengths characterizing its possible interactions with all other clones), which represents a measure of the tolerance threshold of the clone: Clones corresponding to poorly (highly) connected nodes are easily (hardly) allowed to respond to the cognate stimulus. In this way the idiotypic network maintains a regulatory role, where a ''core" of highly (weighted) connected clones acts as a safe-bulk against self-reactions.
%
%
Experimental evidence of this phenomenon has been obtained along the years \cite{varela2,varela3,varela1}, but, even so, given the huge size of the B-repertoire, an extensive experimental exploration has always been out of reach, in such a way that the initial promising perspectives offered by the theory were never robustly actualized,  and interest in this approach diminished.

Conversely, according to the modern two-signal model, the activation of a B-cell (i.e. antibody production and clonal expansion of its lineage) requires two signals in a given (close) time interval: the first one is delivered by the antigen binding to the BCR, the second one is provided
by a helper T lymphocyte, which elicits the B-growth through cytokines \footnote{Cytokines constitute a wide class of cell-signaling protein molecules \cite{chitochine} and among them, e.g. interleukines and interferones, work as immunomodulating agents; the vast majority of these are produced by helper T cells and, as a whole, cytokines are able to both elicit or suppress immune response. For example, Interleukin-2 (IL-2) acts in an autocrine manner to
stimulate B and T cell proliferation, while Interleukin-10 (IL-10) inhibits  responses by reducing MHC expression and the synthesis of eliciting cytokines as IL-2 (or TNF-$\alpha$ and IL-5) \cite{janaway}. The secretion of a certain cytokine, for instance IL$2$ rather than IFN$\gamma$, depends on the inflammatory state and on the concentration of ligands on helper TCR \cite{chitochinebook}.}. In the absence of the second signal, armed clones enter a ''safe mode" \cite{kitamura,anergy3}, being unable to either proliferate or secrete immunoglobulins. This explanation for anergy largely prevailed as, being based on a local mechanism, its experimental evidence is undoubtable, however, it raises the puzzling question of how self-directed B-cells become ''invisible" to helpers \cite{goodnow4} and, also, it does not incorporate previous findings of Varela picture, whose experimental evidences should however be framed in this prevailing scheme.

Aim of this paper is trying to answer these questions through techniques stemmed from theoretical physics: Interestingly, the scenario we outline  robustly evidences that highly connected B cells are transparent to helpers, hence merging the two mechanisms for anergy.

\section{Methods}
In this work we rely on a statistical-mechanics (SM) modellization of the immune system. Indeed, SM, based on solid pillars such as the law of large number  and the maximum entropy principle \cite{MAP1}, aims to figure out collective phenomena, possibly overlooking the details of the interactions to focus on the very key features.
%
%
%
Despite this certainly implies a certain degree of simplification, SM, merging thermodynamics \cite{ellis} and information theory \cite{kinchin1}, has been successfully applied to a wide range of fields, e.g.,  material sciences \cite{allen,frenkel}, sociology \cite{brock1,brock2}, informatics \cite{marc}, economics \cite{ton1,bouchaud}, artificial intelligence \cite{amit,ton2}, and system biology  \cite{enzopnas,kaufman}; SM was also proposed as a candidate instrument for theoretical immunology in the seminal work by Parisi \cite{parisi}.
Indeed, the systemic perspective offered by SM nicely fits emergent properties as collective effects in immunology, as for instance discussed by Germain: ''as one dissects the immune system at finer and finer levels of resolution, there is actually a decreasing predictability in the behavior of any particular unit of function", furthermore, "no individual cell requires two signals (...) rather, the probability that many cells will divide more often is increased by costimulation" \cite{germain}. Understanding this averaged behavior is just the goal of SM.

Moreover, concepts such as ``decision making'', ``learning process'' or ``memory'' are widespread in immunology \cite{chakra, depino,floreano}, and shared by the neural network sub-shell \cite{amit,ton2} of disordered SM \cite{MPV}: Clones, existing as either active or non-active and being able to collectively interact, could replace
the digital processing units (e.g. flip flops in artificial intelligence \cite{AI1}, or neurons in neurobiology \cite{AI2}) and cytokines, bringing both eliciting and suppressive chemical signals, could replace connections (e.g. cables and  inverters in artificial intelligence, or synapses in neurobiology).

As a last remark, we stress that, as typical in SM formalization (see e.g. \cite{ton2}), we first develop the simplest scenario, namely we assume symmetry for the interactions among B and T cells. Despite this is certainly a limit of the actual model, it is has the strong advantage of allowing a clear equilibrium picture still able to capture the phenomenology we focus on, and whose off-equilibrium properties (immediately achievable in the opposite, full asymmetric, limit) should retain strong similarities with the present picture and will be addressed in future investigations.
\newline

Having sketched the underlying philosophy of our work, we highlight our two key results: We first consider the B-H network and show that \textit{helpers are unable to communicate with highly connected B-cells}; Then, we consider the set of B clones and show that a minimal (biased) learning process, during B-cell clonal deletion at ontogenesis, can shape the final repertoire such that \textit{highly connected B clones are typically self-directed}. These two points together allow to merge the two-signal model and Varela's theory.
\newline
The plan of the paper can be summarized by the following syllogism:
\newline
\newline
{\em Part I: Anergy induced by T cells and the "two-signal model".}\\
\begin{itemize}
\item Fact: The response of B-cells is prompted by two signals: the presence of an antigen and the ''consensus'' by an helper T lymphocyte.

\item Consequence: The ensembles made of by B and helper clones interact as a (diluted \cite{PRL}) bilayer restricted Boltzmann machine.

\item Consequence: This system is (thermodynamically) equivalent to an associative ''neural'' network, whose equilibrium states correspond to optimal orchestrations of T cells in such a way that a B clone is maximally signaled and hence prompted to react; each equilibrium state is univocally related to a B clone. Remarkably, the activation of B-clones with high weighted connectivity corresponds to negligible basins of attraction, hence they are rarely signaled by helpers.
\end{itemize}
{\em Part II. Anergy induced by B cells and "Varela theory".}\\
\begin{itemize}
\item Fact: Antibodies (as any other protein) are not random objects (for instance, randomly generated proteins may not even be able to fold into a stable structure \cite{ton}) \cite{bialek}. Hence, once expressed trough e.g. bit-strings of information, the related entropy is not maximal.
\item Consequence: In the idiotypic network where B-clones are nodes and (weighted) links among them mirror the interactions through the related antibodies, nodes with higher weighted connectivity are lazier to react and typically self-directed (Varela Theory).
\end{itemize}

\section{Preliminary remarks on the structure of the B-network}


There are several approaches in estimating the structure, size and shape of the mature B repertoire. For instance, in their pioneering works, Jerne and Burnet used a coarse-grained description in terms of epitopes and paratopes \cite{jerne,burnet}, then Perelson extended (and symmetrized) them introducing a shape space \cite{perelson}, De Boer and coworkers dealt directly with peptides of fixed length \cite{deboer}, while Bialek, Callan and coworkers recently used the genetic alphabet made of by the VDJ genes codifying for the heavy and light chains of the immunoglobulins \cite{bialek} \footnote{Furthermore, a similar approach implies extremely interesting results for TCRs of helper cell lineage too \cite{bialek2}, but anergic signals in T lymphocytes seem more subtle \cite{anergy1,anergy2}, and will not deepened in this paper.}.

Proceeding along a general information theory perspective, we associate to each antibody, labeled as $\mu$, a binary string $\Psi_{\mu}$ of length $L$, which effectively carries information on its structure and on its ability to form complexes with other antibodies or antigens.
Since antibodies secreted by cells belonging to the same clone share the same structure, the same string $\Psi_{\mu}$ is used to encode the specificity of the whole related B clone. In this way, the repertoire will be represented by the set $\mathcal{B}$ of properly generated strings and its cardinality $N_B = |\mathcal{B}|$ is the number of clones present in the system.
%
$L$ must be relatively short with respect to the repertoire size $N_B$, i.e. $L = \gamma \ln N_B$, $\gamma \in \mathbb{R}^+$ \cite{BA1}. This choice stems from both the probabilistic combinatorial usage of the VDJ recombination \cite{bialek} (when thinking at bit-string entries as genes)
and pioneering direct experimental evidence \cite{nobel} (when thinking at bit-string entries as epitopes).


Antibodies can bind each-other through ``lock-and-key'' interactions, that is, interactions are mainly hydrophobic and electrostatic and chemical affinities range over several orders of magnitude \cite{janaway}.
This suggests that the more complementary two structures are and the more likely (on an exponential scale) their binding.  We therefore define $\chi$ as a Hamming distance
\begin{equation} \label{eq:compl}
\chi_{\mu \nu}= \sum_{k=1}^L [\Psi_{\mu}^k(1-\Psi_{\nu}^k) + \Psi_{\nu}^k(1-\Psi_{\mu}^k) ],
\end{equation}
to measure the complementarity between two bit-strings $\Psi_{\mu}, \Psi_{\nu}$, and introduce a phenomenological coupling (whose details will be deepened in Sec.~IV, see also \cite{BA1,PRE})
\be\label{coupling}
J_{\mu \nu} \propto e^{\alpha \chi_{\mu \nu}},
\ee
where $\alpha$ tunes the interaction strength.
In this way, a network where nodes are B-clones, and (weighted) links are given by the coupling matrix $\pmb{J}$,  emerges (see Fig.~$1$, uppermost panel, and \cite{BA1,PRE,PREdeutch,PREdeutch1,PREdeutch2} for details). This formalizes Jerne's idiotypic network.

In general, several links may stem from the same node, say $\mu$, and we define its weighted degree as $W_{\mu}= \sum_{\nu=1}^{N_B} J_{\mu \nu} $.
When the system is at rest, we can argue that all B clones are inactive, so that if clone $\mu$ is stimulated, $W_{\mu}$ can be interpreted as the ``inertia'' of lone $\mu$ to react, due to all other cells \cite{BA2}: This mechanics naturally accounts also for the low dose phenomenon \cite{janaway,BA1,BA2}.


Finally, it is worth considering how $W$ is distributed as this provides information about the occurrence of inertial nodes in the system.
Exploiting the fact that couplings $J_{\mu \nu}$ are log-normally distributed \cite{PRE}, one can approximate the distribution $P(W)$ as
\begin{equation}
P(W) \sim \frac{1}{W \sqrt{2 \pi} \sigma} e^{- \frac{(\log W - \mu)^2}{2 \sigma^2}},
\end{equation}
in such a way that mean and variance read as $E(W)= e^{\mu + \sigma^2/2}$, $V(W)=(e^{\sigma^2}-1)e^{2\mu+\sigma^2}$, respectively (a detailed discussion on the parameters $\sigma$ and $\mu$ can be found in Sec.~V and in Appendix Five).

\begin{figure}\label{disegni}
\begin{center}
\includegraphics[width=7cm]{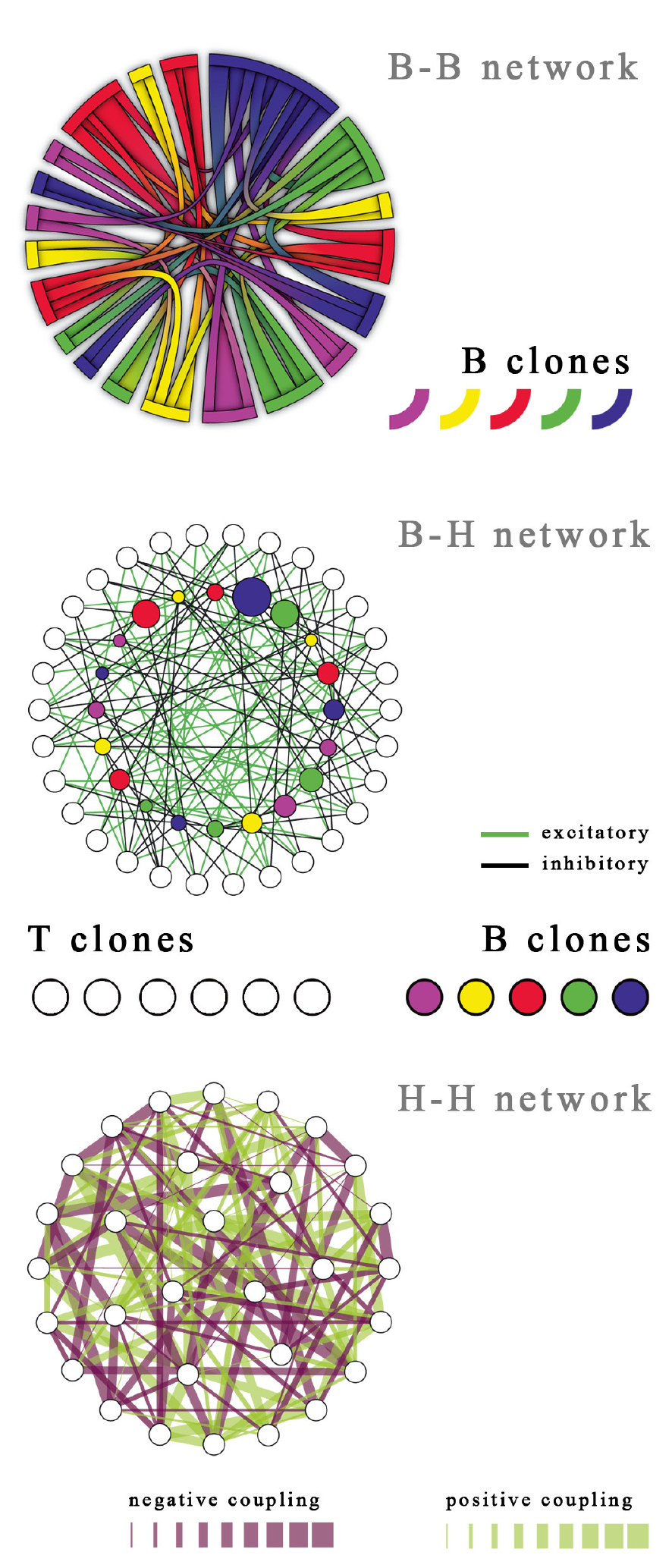}
\caption{Schematic representation of the immune networks considered here, where we fixed $N_H=30$ and $N_B=20$.
The uppermost plot describes the B-B network: each B-cell $\mu$ corresponds to a different arch, whose length is proportional to the related weighted degree $W_{\mu}$, and the interaction between cells $\mu$ and $\nu$ corresponds to the link connecting the related arches, whose thickness is proportional to $J_{\mu \nu}$.
The middle plot describes the bipartite B-H network: the external set of white circles corresponds to the set of T cells, while the internal set of colored circles corresponds to the set of B cells and their size is proportional to the related weighted degree, according to the plot above. The interaction $\xi$ between T cells and B cells can be either excitatory (bright link) or inhibitory (dark link).
The lowermost plot describes the H-H network: the white circles correspond to the set of T cells and connections between them are drawn according to formula $\sum_{\mu} (\xi_i^{\mu}\xi_j^{\mu})/W_{\mu}$, as explained in the text; the color and the thickness of the link carry information about the sign and the magnitude of the coupling, respectively.}
\end{center}
\end{figure}

We stress that the log-normal distribution evidenced here agrees with experimental findings \cite{carneiro1}. Furthermore, its envelope remains log-normal even if the network is under-percolated \cite{PRE}. Thus, in order to have a broad weighted connectivity, the effective presence of a large, connected B-network is not a requisite, but, basically, the mere existence of small-size components, commonly seen in experiments \cite{cazenave,lanzetta}, is needed.

\section{Anergy induced by T cells and the "two-signal model".}

\subsection{Stochastic dynamics for the evolution of clonal size}
We denote with $b_{\mu} \in \mathbb{R}$ the ''degree of activation" of the B clone $\mu$ with respect to a reference value $b_0$, such that if the clone is in its equilibrium (i.e., at rest) $b_{\mu}=b_0$, while if the clone is expanded (suppressed) $b_{\mu}> b_0$ ($b_{\mu} < b_0$); again, we adopt the simplest assumption of fixing a unique reference state $b_0=0$ for all the clones; the case of tunable $b_0$ was treated in \cite{JTB1}.

Concerning T cells, both helper and regulatory sub-classes share information with the B branch via cytokines. Hence, we group them into a unique ensemble of size $N_H$, and denote the state of each clone  by $h_i (i=1,...,N_H)$; hereafter we call them simply ''helpers".
We take  $h_i = \pm 1$ such that $h_i=+1$ stands for an active state (secretion of cytokines) and vice versa for $-1$; actually the choice of binary variables is nor a biological requisite neither a mathematical constraint, but it allows to keep the treatment as simple as possible, yet preserving the qualitative features of the model that we want to highlight.

We define $\epsilon \equiv N_B/N_H$ and, to take advantage of the central limit theorem (CLT), we focus on the infinite volume (thermodynamic limit, TDL), such that, as $N_B\to \infty$ and $N_H\to \infty$, $\epsilon$ is kept constant as, experimentally, the global amount of helpers and of B-clones is comparable.

Recalling that B clones receive two main signals,  i.e. from other B clones and from T ones, we can introduce the Langevin dynamics for their evolution as
\be\label{langevinfull}
\tau \frac{d b_{\mu}}{dt} =
 \sum_{\nu=1}^{N_B} J_{\mu \nu} (b_{\nu}-b_{\mu}) + \sqrt{\frac{\beta}{N_H}}\sum_{i=1}^{N_H} \xi_i^{\mu}h_i + J_{k \mu} A_k + \sqrt{\tau'}\eta_{\mu},
\ee
where $\tau$ rules the characteristic timescale of B cells and $\tau'$ is the timescale of a white noise $\eta \in \mathcal{N}[0,1]$. The ratio between the influence of the noise on the B-H exchanges and the influence on the B-B interactions is tuned by $\beta$.
The coupling between the $\mu$-th B clone and the $i$-th T clone is realized by the ensemble of cytokine $\xi_i^{\mu}$ (see Fig.~$1$, middle panel) and $A_k$ is a generic antigenic peptide that interacts with B-clones through the coupling $J_{k \mu}$.

As far as all the interactions are symmetric \footnote{The assumption of symmetry is the standard first step when trying the statistical-mechanics formulation \cite{amit}. The general case would still retain the same qualitative behavior \cite{ton1}.}, the Langevin dynamics admits a Hamiltonian description as
$$\tau \frac{d b_{\mu}}{dt} = - \frac {d}{d b_{\mu}} \mathcal{H}_{N_H,N_B}(b,h|J, \xi) + \tau' \eta_{\mu},$$
where, by integration over $b_{\mu}$,
\be\label{accadue}\small
\mathcal{H}_{N_H,N_B} = \sum_{\mu, \nu}^{N_B, N_B} \frac{J_{\mu \nu}}{4} (b_{\mu} - b_{\nu})^2 - \sqrt{\frac{\b}{N_H}}\sum_{i, \mu}^{N_H, N_B} \xi_i^{\mu} h_i b_{\mu}  - \sum_{\mu}^{N_B} J_{\mu k } b_{\mu}A_k.
\ee
Each contribution appearing in the r.h.s. of the previous equation is deepened in the following:

\begin{itemize}
\item The first term comes from B-clone interactions via immunoglobulin, which is translated into a diluted ``ferromagnetic'' coupling $J_{\mu \nu} \geq 0$, as B clones tend to ''imitate" one another.
Notice that the square $(b_{\mu} - b_{\nu})^2$ generalizes the ferromagnetic behavior, typically referred to binary Ising spins, to the case of ``soft spins'' variables: the usual, two-body term $b_{\mu}b_{\nu}$ \cite{ellis} is clearly recovered, while the two extra terms $b_{\mu}^2$ encode a one-body interaction that here promotes B-cell quiescence in the absence of stimulation.

\item The second term represents the coupling between B and T clones, mediated by cytokines: The cytochine $\xi_i^{\mu}$ is meant to connect cells of the $i$-th helper clone and those of the $\mu$-th B one. The message conceived can be either excitatory ($\xi_i^{\mu} =+1$, e.g. an eliciting  Interleukin-$2$) or inhibitory ($\xi_i^{\mu} =-1$, e.g. a suppressive Interleukin-$10$) and here is assumed to be a quenched variable, such that the one with inhibitory effects can be associated to a regulatory cell and, viceversa, the one with stimulating effect to an helper cell.
Note that the choice $\pm 1$ for $\xi_i^{\mu}$ is only a convenient requisite encoding two opposite effects, while, clearly, their world is by far richer \cite{chitochinebook}, and, in principle, also mathematically accessible.

\item The third term  mimics the interaction of the generic $b_{\mu}$ clone with the antigenic peptide $A_{k}$
where $J_{\mu k }$ encodes their coupling strength and can be defined according to eq.~\ref{coupling}.
\end{itemize}

Interestingly, in the Hamiltonian \ref{accadue}, the first term recovers Jerne's idiotypic network theory, the second one captures the two-signal model and the third one recovers Burnet's clonal selection theory: within this SM framework the three approaches are not conflicting, but, rather, interplaying.

Close to equilibrium, whose investigation is our first goal, the antigenic load is vanishing ($A_k=0$ for all $k$) and the anti-antibodies can consequently be neglected ($b_{\mu} b_{\nu} \sim 0$), hence
the Langevin process defined in eq. \ref{langevinfull} simplifies to
\begin{equation} \label{eq:langevinshort}
\tau \frac{d b_{\mu}}{dt} = - \frac{d}{db_{\mu}}\left( \frac{1}{2}\sum_{\nu}W_{\nu}b_{\nu}^2- \sqrt{\frac{\beta}{N_H}}\sum_{\nu,i}^{N_B,N_H}\xi_i^{\nu}h_i b_{\nu} \right) + \sqrt{\tau'}\eta_{\mu},
\end{equation}
where $W_{\mu}=\sum_{\nu=1}^{N_B} J_{\mu \nu}$  is the (weighted) connectivity of the $\mu$-th node (clone) of the B-network.
\newline
Therefore, the Hamiltonian of the process is
\be\label{ammazza}
\hat{\mathcal{H}}_{N_H,N_B}=\frac{1}{2}\sum_{\nu}^{N_B}W_{\nu}b_{\nu}^2 - \frac{\beta}{N_H}\sum_{i,\nu}^{N_H,N_B}\xi_i^{\nu}h_i b_{\nu},
\ee
and its properties will be addressed in the next section through statistical mechanics.

\subsection{The equivalence with associative networks}
Once the effective Hamiltonian is defined through eq.~\ref{eq:langevinshort},
the classical statistical mechanics package can be introduced; this implies the partition function
\be\label{zeta}
Z_{N_H, N_B}(\beta|\xi,W) =  \sum_{\{ h \}} \int \prod_{\mu}^{N_B} d b_{\mu} e^{- \frac{\sum_{\mu}^BW_{\mu}b_{\mu}^2}{2} + \sqrt{\frac{\beta}{N_H}} \sum_{\mu, i}\xi_i^{\mu}h_i b_{\mu}},
\ee
and the quenched free-energy (neglecting constant terms which do not affect the scenario)
\begin{equation} \label{eq:free}
A(\beta,\epsilon|P(W)) = \lim_{N_H, N_B \to \infty}\frac1{N_H} \mathbb{E}\ln Z_{N_H,N_B}(\beta|\xi,W),
\end{equation}
where $\mathbb{E}$ averages over both the $\xi$ and the $W$ distributions.
\newline
Notice that the idiotypic contribution in the stochastic process (\ref{eq:langevinshort}) implicitly  generates a  Gaussian distribution for the activity of the B-clones
\begin{equation} \label{eq:P_b}
P(b_{\mu}|W) \propto \exp\left(-W_{\mu}b_{\mu}^2/2 \right),
\end{equation}
which ensures convergence of the Gaussian integrals.
This is consistent with commonly observed data and ensures convergence of the integrals in the partition function \ref{zeta}; interestingly, $W_{\mu}^{-1}$ plays as variance.
%
\newline
A crucial point is that the integrals over $\{ b_{\mu}\}$ in the partition function \ref{zeta} can be calculated explicitly to give
\be \label{eq:Z}
Z_{N_H,N_B}(\beta|\xi,W) =  \sum_{\{ h \}} \exp \left( \frac{\beta}{2N_H}\sum_{i,j}^{N_H,N_H}\sum_{\mu}^{N_B} \, \frac{\xi_i^{\mu}\xi_j^{\mu}}{W_{\mu}} \, h_i h_j \right).
\ee
The previous expression deserves attention because it corresponds to the partition function of a (log-normally weighted)  Hopfield model for neural networks (\cite{amit}), (see Fig.~$1$, lowest panel):
Its Hebbian kernel suggests that the network of helpers is able to orchestrate strategies (thought of as patterns of cytokines) if the ratio $\epsilon = N_B/N_H$ does not exceed a threshold \cite{JTB1}, in agreement with the breakdown of immuno-surveillance occurring whenever the amount of helpers is too small (e.g. in HIV infections) or the amount of B is too high (e.g. in strong EBV infections) \footnote{These capabilities of the system are minimally focused in this paper and again we refer to \cite{JTB1} for further insights and to \cite{JTB2,PRL} for the investigation of its parallel processing performances (namely the ability of managing several clones simultaneously).}.

\begin{figure}\label{disegni}
\begin{center}
\includegraphics[width=5.5cm]{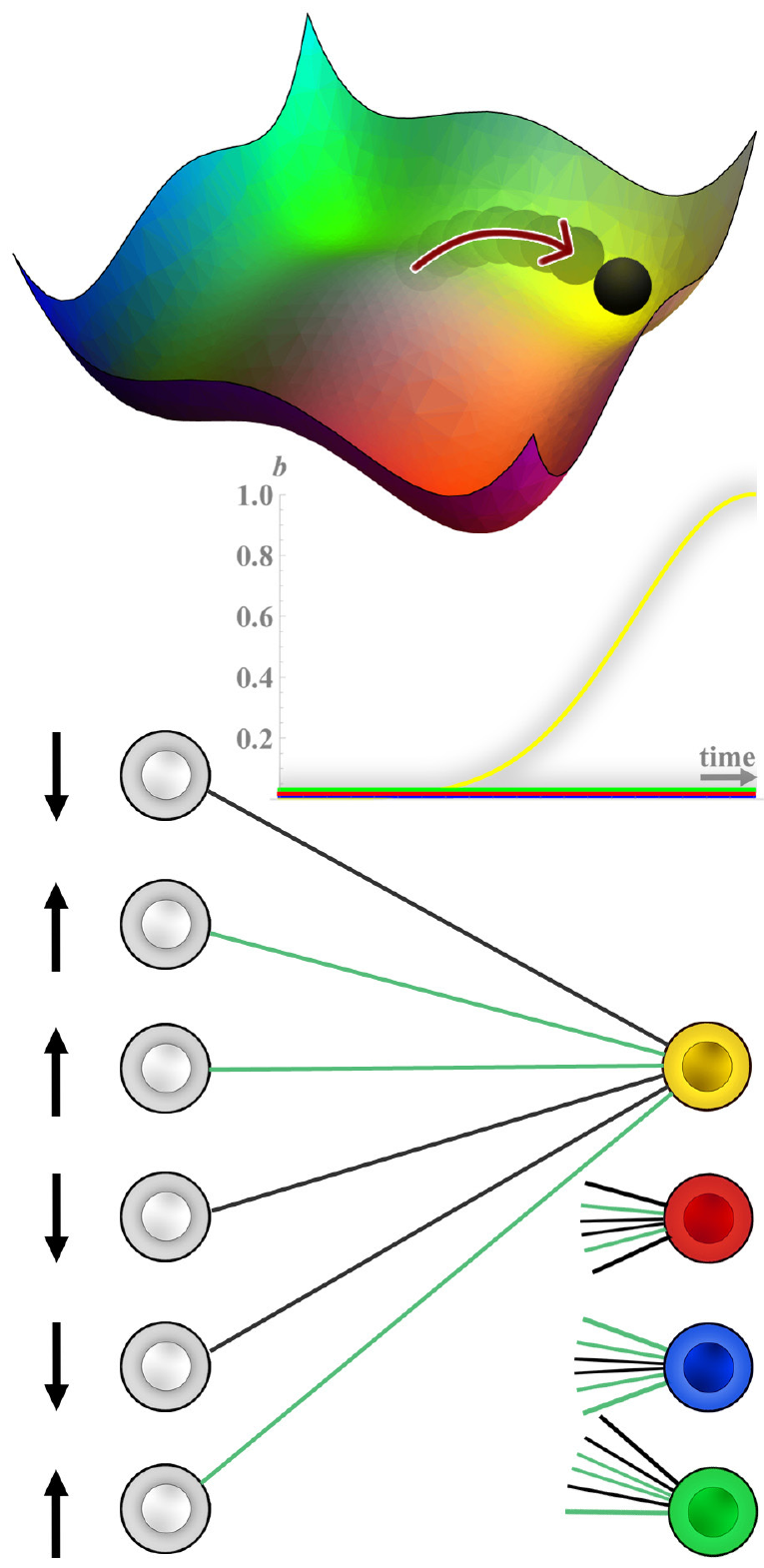}
\caption{Schematic representation of the consequence
of retrieval capabilities by the helper network in the
bipartite network made of by both helpers and B-clones: In the upper panel a free-energy landscape of the helper network, with four minima (each corresponding to retrieval of instruction for a particular B-clone) is shown. The black ball represents the state of the system, which is driven into the yellow hole (e.g. due to antigenic stimulation). Consequently, as all the helpers in the bipartite network (lower panel) become parallel to the sign of the cytokines linking them to the yellow B-clone. This results in maximal strength conferred to the retrieved clone, that undergoes clonal expansion. The latter is represented in the middle plot, together with the lack of growth by the other clones (not-retrieved).}
\end{center}
\end{figure}

\subsection{High connectivity leads to anergy}
As anticipated, the network made of by helper cells can work as a neural network able to retrieve ``patterns of information''. There are overall $N_B$ patterns of information encoded by cytokine arrangement $\{ \xi \}$ and the retrieval of the pattern $\mu$ means that the state of any arbitrary $i$-th T clone agrees with the cytokine $\xi_i^{\mu}$, namely $h_i \xi_i^{\mu} = +1$; this ultimately means that clone $\mu$ is maximally stimulated.
A schematic representation of retrieval performed by T cells and of its consequence on the repertoire of B cells is depicted in Fig.~$2$.

Here, with respect to standard Hopfield networks, Hebbian couplings are softened by the weighted connectivity $W_{\mu}$ and this has some deep effects.
In fact, the patterns of information which can be better retrieved (i.e. the clones which can be more intensively signaled) are those corresponding to a larger signal, that is, a smaller $W$. Thus, B-clones with high weighted connectivity (the safe-bulk) can not be effectively targeted and, in the TDL, those B-clones exhibiting $W \rightarrow \infty$ are completely ``transparent" to helper signaling.

Deepening this point is now mainly technical. We introduce the $N_B$ pattern-overlaps $\langle m_{\mu} \rangle$, which measure the extent of pattern retrieval, i.e. signaling on clone $\mu$, and are defined as
$ \langle m_{\mu} \rangle = \mathbb{E} N_H^{-1} \Omega(\sum_i^{N_H} \xi_i^{\mu}h_i)$,
where $\Omega$ is the standard Boltzmann state \cite{ellis} associated to the free energy \ref{eq:free}, which allow to rewrite the Hamiltonian corresponding to Eq.~(\ref{eq:Z}) as
\be\small
\mathcal{H}_{N_H,N_B}(h|\xi,W) = \frac{-1}{2 N_H}\sum_{i, j}^{N_H, N_H} (\sum_{\mu}^{N_B} \frac{ \xi_i^{\mu}\xi_j^{\mu}}{W_{\mu}} )h_i h_j = -\frac{N_H}{2}\sum_{\mu}^{N_B} \frac{m_{\mu}^2}{W_{\mu}}.
\ee
Now, free energy minimization implies that the system spontaneously tries to reach a retrieval state where $\langle m_{\mu} \rangle \to 1$ for some $\mu$. Of course, this is more likely for clones $\mu$ with smaller $W_{\mu}$, while highly connected ones  are expected not to be signaled (pathological cases apart, i.e. no noise $\beta \to \infty$, or giant clonal expansions $b_0 \to \infty$ limits).

Note that $\langle m \rangle_{\mu}=1$ (gauge-invariance apart) means that all the helpers belonging to the clone $i$
are parallel  to their corresponding cytokine, hence  if $\xi_i^{\mu}$ is an eliciting messenger,
the corresponding helper $h_i$ will be firing, viceversa for $\xi_j^{\mu}=1$ the corresponding
helper $h_j$ will be quiescent, so to confer to the $b_{\mu}$ clone the maximal expansion field.
\newline
In order to figure out the concrete existence of this retrieval, we solved the model through standard replica trick \cite{MPV}, at the replica symmetric level (see Appendix One), and integrated numerically the obtained self-consistence equations, which read off as
\begin{eqnarray}\nonumber
\langle m_{1}(\epsilon,\beta) \rangle &=&  \langle \langle \xi^{1}\tanh\Big( \beta ( m_1 \xi^1 / W_1 + \sqrt{\epsilon r}z )\Big)\rangle_{z}\rangle_{\xi,W},\\ \nonumber
\langle q (\epsilon,\beta)  \rangle &=&  \langle \langle \tanh^2\Big( \beta ( m_1 \xi^1 / W_1 + \sqrt{\epsilon r}z )\Big)\rangle_{z}\rangle_{\xi,W}, \\
\langle r (\epsilon,\beta)  \rangle &=& \lim_{N_H \to \infty} \frac{1}{\epsilon  N_H}\sum_{\mu>1}^{N_B}\frac{q}{[W_{\mu}-\beta(1-q)]^2}.
\end{eqnarray}
In this set of equations, we used the label $1$ to denote a test B-clone $\mu=1$, which can be either a self node (i.e. with a high value of $W_1$, infinite in the TDL) or a non-self one (i.e. with a small value of $W_1$, zero in the TDL). While the first equation defines the capability of retrieval by the immune network as earlier explained, $q$ is the Edward-Anderson spin glass  order parameter \cite{MPV} and $r$ accounts for the slow noise in the network due both to the number of stored strategies and to the weighted connectivity \footnote{These equations generalize the Hopfield equations \cite{amit} which clearly are recovered by setting $W_{\mu}=1$ for all $\mu=1,..,B$.}.

As shown in the Appendix Two, the equations above can be solved in complete generality. Here, for simplicity, we describe the outcome obtained by replacing
all $W_{\mu}$ with $\mu \neq 1$ (as $\mu=1$ is the test-case) with their average behavior, namely $\langle W \rangle = \int dW P(W) W$; this assumption makes the evaluation of the order parameter $r$ much easier, yet preserving the qualitative outcome.
\begin{figure*}\label{disegni_W}
\begin{center}
\includegraphics[width=14cm]{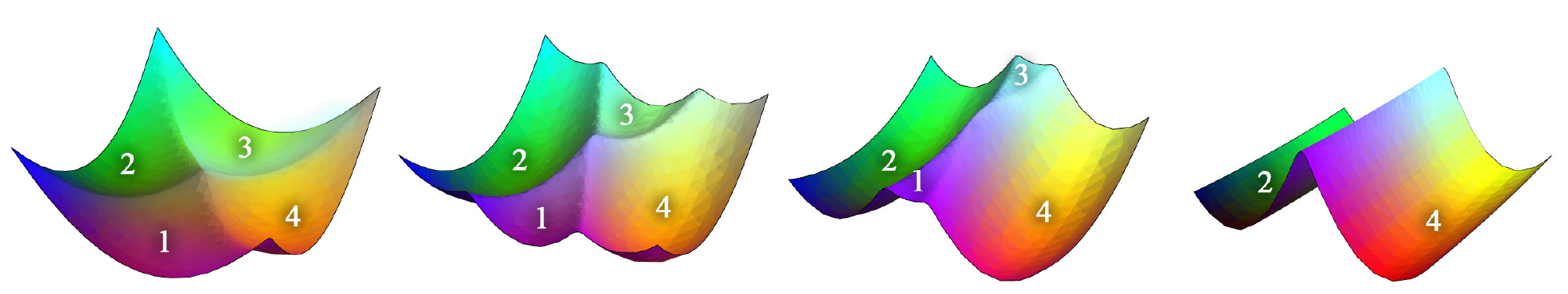}
\caption{Schematic representation of the (free-energy) basins of attractions for a toy-system starting (at left) with four minima (hence four retrievable patterns). Each minimum contains information addressed to the corresponding $B$-clone so that four B-clones $B_1,B_2,B_3,B_4$ can be instructed in the initial configuration. From left to right we fix $W_2=W_4=1$ always, while we increase progressively $W_1=W_3=1,5,10,100$ (and we show the resulting basins of attraction from left to right). Note that at value of the weighted connectivity $W_1=W_3 = 100$, the corresponding minima completely disappear hence instructions to the corresponding $B$-clones (which are broadly interacting as their $W$ is much higher than $W_2=W_4=1$) can not be supplied by helpers.}
\end{center}
\end{figure*}

We now focus on the two limiting cases: $W_1 << \langle W \rangle$, which accounts for a non-self node, and  $W_1 >> \langle W \rangle$, which mirrors the self counterpart.
\newline
In the former case, the slow noise is small (vanishing as $\langle W \rangle \to \infty$), consequently, the non-self nodes live in a free environment and the corresponding equations for their retrieval collapse to the not-saturated Hopfield model \cite{amit}. Hence, retrieval should be always possible (ergodic limit apart), therefore, in this case, helpers can effectively signal clone $1$.

Conversely, in the latter case, namely dealing with a self-node, it is straightforward to check that the noise rescaling due to $W$ implies a critical noise level for the retrieval $\beta^{-1} \sim W_1^{-1} \sim 0$ (as $W_1$ is ideally diverging in the thermodynamic limit, see Fig.~$5$ and Sec.~V). As a result, under normal conditions, the retrieval of patterns enhancing self-node clonal expansions is never performed by helpers: This behavior mimics anergy as a natural emergent property of these networks.
\newline
As a further numerical check we performed Monte Carlo simulations which are in agreement with these findings.

\section{Anergy induced by B cells and "Varela theory".}

So far we showed that helper cells are unable to exchange signals with highly connected B-clones, however, the reason why the latter should be self-directed is still puzzling. Now, we build a basic model for the ontogenetic process of B cells, which solely assumes that self proteins are not random objects, and we show that survival clones expressing large self-avidity are those highly connected.

\subsection{Ontogenesis and the emergence of a biased repertoire.}

During ontogenesis in the bone marrow, B-cell survival requires sufficiently strong binding to at least one self molecule (positive selection), but those cells which bind too strongly are as well deleted (negative selection): such conditions ensure that surviving B cells are neither aberrant nor potentially harmful to the host \cite{kosmir1,kosmir2}.

To simulate this process, we model the ensemble of self-molecules as a set $\mathcal{S}$ of strings $\Phi_{\mu}$, of length $L$, whose entries are extracted independently via a proper distribution. The overall number of self-molecules is $|\mathcal{S}| = N_S$, that is, $\mu = 1, ..., N_S$.

As stated in the introduction, despite a certain degree of stochasticity seems to be present even in biological systems, proteins are clearly non-completely random objects \cite{ton}: Indeed, the estimated size of the set of self-proteins is much smaller than the one expected from randomly generated sets \cite{bialek}. Within an information theory context, this means that the entropy of such repertoire is not maximal, that is, within the set $\mathcal{S}$ some self-proteins are more likely than others (see Appendix Three).

In order to account for this feature, we generate $\mathcal{S}$ extracting each string entry $i$ according to the simplest biased-distribution
\begin{equation} \label{eq:P_self}
P_{\mathrm{self}}(\Phi_i^{\mu} | \bar{a})= \delta(\Phi_i^{\mu}-1) \frac{1+\bar{a}}{2} + \delta(\Phi_i^{\mu}-1) \frac{1-\bar{a}}{2},
\end{equation}
where $\delta(x)$ is the Kronocker delta and $\bar{a} \in [-1,1]$ is a parameter tuning the degree of bias, i.e. the likelihood of repetitions among string-bits.
Of course, when $\bar{a}=0$ the complete random scenario is recovered.
We stress that here, looking for minimal requisites, we neglect correlations among string entries \cite{bialek}, in favor of a simple mean-field approach where entries are identically and independently generated.

As underlined above, a newborn B cell, represented by an arbitrary string $\Psi$, undergoes a screening process and the condition for survival can be restated as
\begin{equation} \label{eq:vincoli}
\chi_P < \max_{ \Phi \in \mathcal{S}} \{ \chi(\Psi, \Phi)\} < \chi_N,
\end{equation}
being $\chi_P$ and $\chi_N$ the thresholds corresponding to positive and negative selection, respectively.

As explained in Appendix Four, the value of the parameters $\chi_P$ and $\chi_N$ can be fixed according to indirect measurements, such as the survival probability of new-born B cells: it is widely accepted that human bone marrow produces daily
$\sim 10^7$ B cells, but only $\sim 10^6$ are allowed to circulate
in the body, the remaining $90\%$ undergo apoptosis since targeted as self-reactive ones \cite{science,NB,survival1,survival2}; therefore the expected survival probability for a new-born B cell is $P_{\textrm{surv}} = 0.1$ (see Fig.~$4$, left panel).

Thus, we extract randomly and independently a string $\Psi$ and we check whether Eq.~\ref{eq:vincoli} is fulfilled; if so, the string is selected to make up the repertoire $\mathcal{B}$. We proceed sequentially in this way until the prescribed size $N_B$ is attained (see Appendix Four for more details).

The final repertoire is then analyzed finding that the occurrence of strings entries is not completely random, but is compatible with a biased distribution such as
\begin{equation} \label{eq:P_rep}
P_{\mathrm{rep}}(\Psi_i^{\mu} | a)= \delta(\Psi_i^{\mu}-1) \frac{1+a}{2} + \delta(\Psi_i^{\mu}-1) \frac{1-a}{2},
\end{equation}
where $a$ turns out to be correlated with $\bar{a}$. More precisely, positive values of $\bar{a}$ yield a biased mature repertoire with $a>0$ (see Fig.~$4$, central panel).
Consequently, in the set $\mathcal{B}$ generated in this way, nodes with large $W_{\mu}$, and therefore dissimilar with respect to the average string, are likely to display large affinity with the self repertoire.
To corroborate this fact we measured the correlation $\rho$ between the weighted degree $W_{\mu}$ of a node and the affinity $\max_{ \Phi \in \mathcal{S}} \{ \chi(\Psi_{\mu}, \Phi)\}$ with the self-repertoire finding a positive correlation (see Fig.~$4$, right panel).
We also checked the response of the B-repertoire when antigens are presented, finding that, when a string $\Phi_{\nu} \in \mathcal{S}$ is taken as antigen, the best-matching node, displaying large $W$, needs an (exponentially) stronger signal on BCR in order to react.

Such results mirror Varela's theory \cite{varela3,varela1}, according to which ``self-directed" nodes display a high (weighted) connectivity, which, in turn, induces inhibition.

\begin{figure} \label{fig:trittico}
\includegraphics[width=8.5cm]{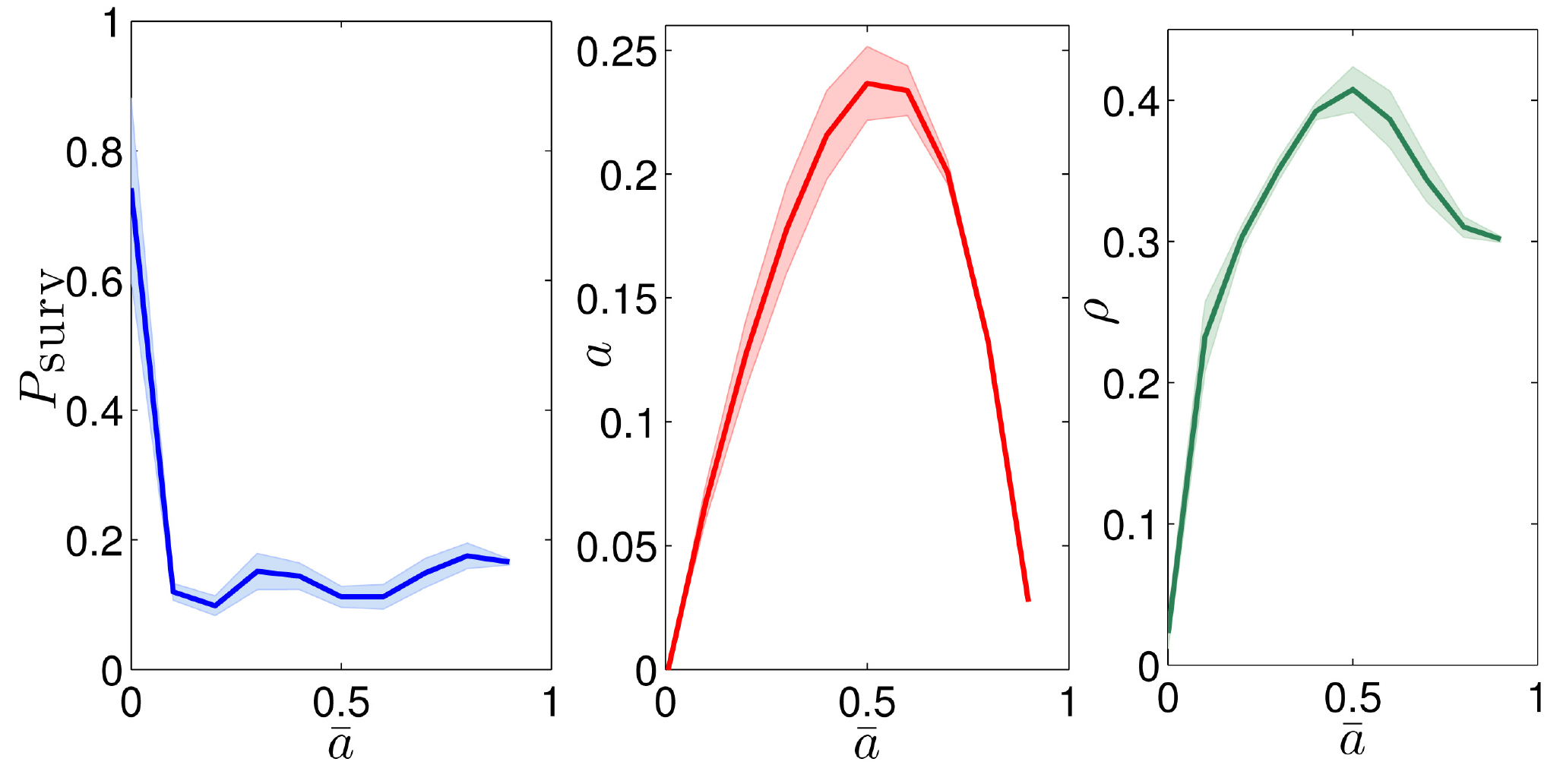}
\caption{Plots from simulations where we generated random strings $\Psi$ and we compared them with those in $\mathcal{S}$ and generated according to the distribution in Eq.~\ref{eq:P_self}. Strings $\Psi$ fulfilling the condition $\ref{eq:vincoli}$ are retained and the survival probability $P_{\textrm{surv}}$ is measured and plotted versus $\bar{a}$ (left panel). The final repertoire $\mathcal{B}$ turns out to be as well biased with degree $a$ depending on $\bar{a}$ (central panel). Moreover, we measured the Spearman correlation coefficient $\rho$, averaged over $\mathcal{B}$, between $W_{\mu}$ and $\max_{ \Phi \in \mathcal{S}} \{ \chi(\Psi_{\mu}, \Phi)\}$ (right panel): notice that a positive value denotes the existence of correlation and gives strong numerical evidence of Varela theory. Data represented in these plots refer to a system where we fixed the size of the B-repertoire $N_B=10^5$ and $\gamma=2, c= 0.5, \Delta=0.4, \chi_P=0.6L$ (see Appendix Four for more details); data were averaged over $10^3$ realizations.}
\end{figure}

Finally, it is worth underlying that, by taking a biased distribution for string entries (i.e., $a \neq 0$), the distribution $P(W)$ for weights occurring in the idiotypic network still retains its logarithmic shape, namely
\begin{equation}
P(W) = \frac{1}{W \sqrt{2 \pi} \sigma} e^{- \frac{(\log W - \mu)^2}{2 \sigma^2}},
\end{equation}
with
\begin{eqnarray}
\mu &=& \log \left[ \frac{ N_B \langle J \rangle_a^2 }{ \sqrt{ \langle J \rangle_a^2 +  (\langle J^2 \rangle_a - \langle J \rangle_a^2)/N_B}}  \right],\\
\sigma^2 &=&\log \left[ \frac{  \langle J^2 \rangle_a - N_B \langle J \rangle_a^2 }{N_B \langle J \rangle_a^2 } +2 \right],
\end{eqnarray}
where $\langle J\rangle_a$ and $\langle J^2 \rangle_a$ are, respectively, the mean value and the mean squared value of coupling $J_{\mu \nu}$ defined in Eq.~(2).
A detailed derivation of these values can be found in Appendix Five, while here we simply notice that, by properly tuning $a$ and $\alpha$, one can recover, in the thermodynamic limit, different regimes characterized by different behaviors (finite, vanishing or diverging) for the average $E(J) \equiv \langle J \rangle_a$ and the variance $V(J) \equiv \langle J^2\rangle_a - \langle J \rangle_a^2$, respectively, as reported in Fig.~$5$.

\begin{figure} \label{fig:esponenti}
\includegraphics[width=8cm]{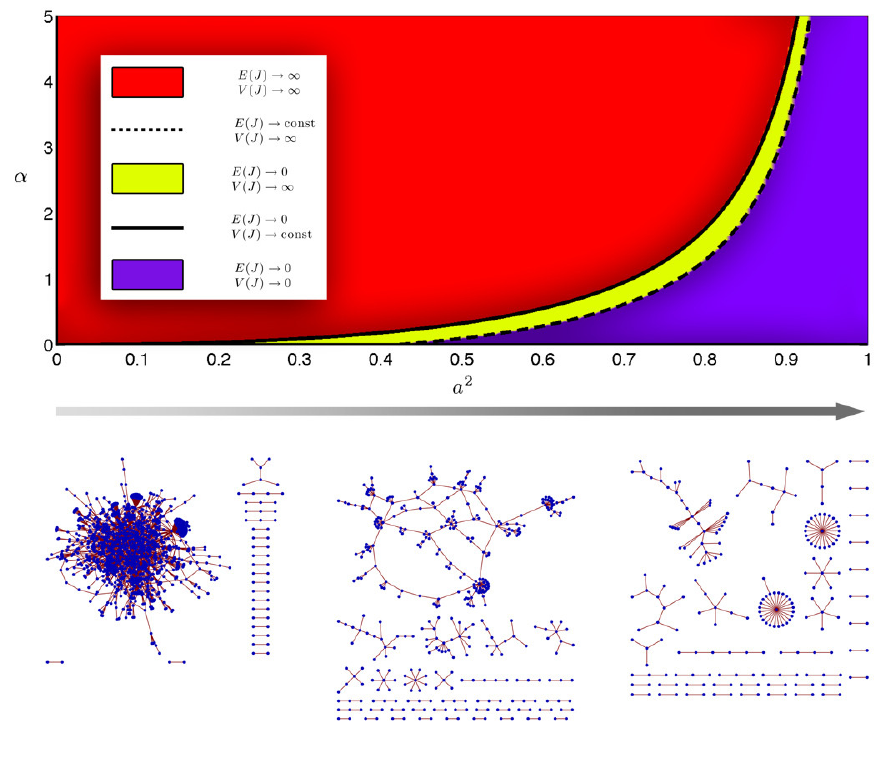}
\caption{In the upper part of this figure we show a phase diagram concerning the distribution of the interaction strengths of the idiotypic network. More precisely, being $\gamma=2$ fixed, we highlight different regions of the $(a^2,\alpha)$ plane, each corresponding to a different behavior of the average coupling $E(J) = \langle J \rangle_a$ and of the variance $V(J) = \langle J^2 \rangle_a - \langle J \rangle_a^2$, as explained by legend.
Different behaviors of $E(J)$ and $V(J)$ can be recast into different topological regimes as envisaged by the graphs depicted in the lower part of the figure, representing particular realizations of the idiotypic network, and referring to the particular choice $\alpha=0.7, N=10^4$ and to different values of $a$ (see also \cite{PRE}).
We underline that different regions imply different thermodynamic regimes which can be associated to different immunological capabilities.}
\end{figure}

\section{Conclusions and outlooks}

In this paper we tried a systemic approach for modeling a subset of the adaptive response of the immune system by means of statistical mechanics.
We focused on the emergent properties of the interacting lymphocytes starting from minimal assumptions on their local exchanges and, as a fine test, we searched for the emergence of subtle possible features as the anergy shown by self-directed B-cells.

First, we reviewed and framed into a statistical mechanics description, the two main strands for its explanation, i.e.  the two-signal model and the idiotypic network. To this task we described the mutual interaction between B cells and (helper and suppressor) T cells as a bi-partite spin glass, and we showed its thermodynamical equivalence to an associative network made of by T cells (helpers and suppressors) alone. Then, the latter have been shown to  properly orchestrate the response of B cells as long as their connection within the bulk of the idiotypic network is rather small.
In the second part we adopted an information theory perspective to infer that highly-connected B clones are typically self-directed as a natural consequence of learning during ontogenic learning.

By merging these results we get that helpers are always able to signal non-self B lymphocytes, in such a way that the latter can activate, proliferate and produce antibodies to fight against non-self antigens. On the other hand, self lymphocytes, due to their large connectivity within the idiotypic network, do not feel the signal sent by helpers.

Therefore, a robust and unified framework where the two  approaches act synergically is achieved. Interestingly, this picture ultimately stems from a  biased learning process at ontogenesis and offers, as a sideline, even a theoretical backbone to Varela theory. We stress that, while certainly the Jerne interactions among B cells act as a key ingredient (and the existence of anti-antibodies or small reticular motifs has been largely documented), an over-percolated B network is not actually required as the distribution of the weighted clonal connectivity remains broad even for extremely diluted regimes.

Furthermore, we note that, within our approach,  while Varela theory is reabsorbed into the two-signal model, the the mutual is not true as clearly other cells (beyond highly connected ones in the B-repertoire), trough other paths, may lack helper signalling and become anergic, hence the two-signal is not necessarily reabsorbed into Varela theory.

Furthermore, the model developed is able to reproduce several other aspects of real immune networks such as the breakdown of immuno-surveillance by unbalancing the leukocitary formula, the low-dose tolerance phenomenon, the link between lymphocytosis and autoimmunity (as for instance well documented in the case of A.L.P.S.\cite{JTB1}) and the capability of the system to simultaneously cope several antigen \cite{JTB2,PRL}.


Despite these achievements, several assumptions underlying this minimal model could be relaxed or improved in future developments, ranging from the symmetry of the interactions, to the fully connected topology of the B-H interactions.

\section{Appendices}

\subsection{Appendix One: The replica trick calculation for the free energy}

In this section we want to figure out the expression of the free energy relative to the partition function (eq. \ref{eq:Z}) of a weighted Hopfield model near saturation (for values $\epsilon \neq 0$) whose weight are drawn accordingly $P(W)$. Its derivation is obtained using the ''replica trick", namely
$$
\log Z = \lim_{n \to 0}\frac{Z^n-1}{n},
$$
within the replica symmetric approximation \cite{MPV}. Through the latter, the free energy $A(\b,\epsilon)$ (hereafter simply $A$ for the sake of simplicity) can be written as

\be\nonumber\small 
A = \lim_{N_H \to \infty} \lim_{n \to 0}\frac{1}{N_H n}\log \langle \sum_{h^1,...,h^n}\exp \big\{ -\beta \sum_{a=1}^n \hi(h^a, \xi) \big\} \rangle_{\xi}
\ee
where we introduced the symbol $a \in (1,\dots,n)$ to label the different replicas and $\langle \cdot \rangle_\xi$ indicates a \emph{quenched average} on the patterns $\xi$. The replicated partition function averaged over the patterns $\xi$ hence reads as
\be
\langle Z_{N_H,N_B}^n(\beta |\xi,W) \rangle_{\xi} =  \bigl \langle \sum_{\{h\}} \exp\Big\{ \frac{\beta}{2N_H}\sum_{\mu,a}^{N_B,n}\frac{(\sum_{i}^{N_H}\xi_{i}^{\mu}h_{i}^a)^{2}}{W_{\mu}} \Big\} \bigr \rangle_{\xi}
\ee
which is equivalent to eq. \ref{eq:Z}. Now, without loss of generality, we suppose to retrieve a number $s$ of memorized patterns and  we divide the sum over the $N_B$ patterns in two sets: the former refers to the retrieved patterns (labeled with the index $\nu=1,\dots,s$) while the latter refers to the not-retrieved ones (labeled with $\mu=s+1,\dots,N_B$).

The retrieved patterns sum can be manipulated introducing $n \times s$ Gaussian variables in order to linearize the quadratic term in the exponent

\begin{eqnarray}\nonumber
    e^{\frac{\beta}{2N_H}\sum_{\mu<s}\sum_a\frac{(\sum_i^{N_H}\xi_i^{\mu}h_i^a)^2}{W_{\mu}} }  &=&   \Bigl\langle \int dm \exp \Big[ -\frac{\beta}{2}\sum_{\mu,a}^{N_B,s}\frac{(m_{\mu}^a)^2}{W_{\mu}}  \\
     &+&  \beta \sum_{\mu,a}^{s,n} \frac{m_{\mu}^{a}}{W_{\mu}} \sum_{i}^{N_H}\xi_{i}^{\mu}h_{i}^a\Big] \Bigr\rangle_{\xi}.
\end{eqnarray}

On the other side, the term corresponding to non retrieved patterns can be written, after some computations including averaging over $\xi$, as

\be \label{ZNR}
e^{\frac{\beta}{2N_H}\sum_{\mu>s}\sum_a\frac{(\sum_i^{N_H}\xi_i^{\mu}h_i^a)^2}{W_{\mu}} } =  \exp{\Big\{ -\frac{1}{2}\sum_{\mu=s+1}^{N_B} \mathrm{Tr} \ln[K(W_\mu)] \Big\}},
\ee
where $K_{ab}(W_\mu) = \delta_{ab}- (\beta/N_H) \sum_{i}^{N_H}  \frac{h_{i}^a h_{i}^b}{W_\mu}$. The previous expression motivates the introduction of the family of $n(n-1)$ order parameters $q_{ab}=\frac 1 {N_H}\sum_i^{N_H} h_{i}^a h_{i}^b$ and their conjugates $\hat{q}_{ab}$ through the identity
\begin{eqnarray}
 \mathbf{1} &=& \prod_{a,b} \int dq_{ab} \ \delta(q_{ab}-\frac 1 {N_H}\sum_{i}^{N_H} h_{i}^a h_{i}^b )    \\\nonumber
 &=&\int \big(\prod_{a,b} dq_{ab}d\hat{q}_{ab}\big) \emph{e}^{i \hat{q}_{ab}(N_H q_{ab}- \sum_{i}^{N_H} h_{i}^a h_{i}^b )}.
\end{eqnarray}
Putting all together and omitting negligible terms in $N_H$, we get
{\small
\begin{align*} \label{}
A(\b,\epsilon &)  = \lim_{N_H \to \infty} \lim_{n \to 0}\frac{1}{N_H n}\log  \int  dm \big(\prod_{a,b} dq_{ab}d\hat{q}_{ab}\big) \\
&\exp\bigg\{ N_H  \bigg[\, -\frac{1}{2 N_H}\sum_{\mu=s+1}^{N_B} Tr \ln \big[ K(W,\{q_{ab}\})\big] \\
& + i \sum_{a,b}^n \hat{q}_{ab}q_{ab}  - \frac{\beta}{2} \sum_{\nu,a}^{s,n} \frac{(m_{\nu}^{a})^{2}}{W_{\nu}}   \\
& +  \langle \ln \sum_{\{h\}} \exp \Big\{ -i \sum_{a,b}^n \hat{q}_{ab} h^a h^b + \beta \sum_{\nu,a}^{N_B,n} \frac{m_{\nu}^{a}}{W_{\nu}} \xi^{\nu} h^a \Big\}  \rangle_\xi \bigg] \bigg\}.
\end{align*}
}

In the last expression, the principal dependence from the system size $N_H$ is in the global factor into the exponent, hence we can obtain the replicated free energy using the saddle-point method, i.e. extremizing the function in the exponent. Under replica-symmetry assumption we get
{\small
\begin{equation}\label{freeEn} 
\begin{split}
& A(m,q,r| \beta,  \epsilon) = -\frac{\beta}{2}\sum_{\nu=1}^s \frac{\langle m_\nu \rangle_{\xi}^2}{W_\nu} \\
& - \lim_{N_H \to \infty} \frac{1}{2 N_H} \sum_{\mu>s}^{N_B} \Big[ \ln\big(1-\frac{\beta}{W_\mu}(1-q)\big) - \frac{\frac{\beta q}{W_\mu}}{1-\frac{\beta}{W_\mu}(1-q)}  \Big] \\
   &  - \frac{\epsilon \beta^2 \langle r \rangle_{\xi}}{2}(1 - \langle q \rangle_{\xi}) + \langle \ln 2\cosh\Big[\beta\Big(\sum_{\nu=1}^{s}\frac{m_{\nu}}{W_{\nu}}\xi^{\nu}+\sqrt{\epsilon r}z\Big)\Big]\rangle_{\xi,\, z}
\end{split}
\end{equation}
}where $\langle \cdot \rangle_z$ indicates the average over the measure $d\mu(z)= \exp{(-z^2/2)}$.
We then obtain the self-consistent equations reported in the main text by extremizing  $A(m,q,r| \beta,  \epsilon)$ with respect to $m, q, r$.

\subsection{Appendix Two: Quenched evaluation of the slow noise order parameter $r$}
As we hinted in the main text and in the previous section, extremizing the free energy \ref{freeEn} with respect to $m$ and $r$ allows to get the self-consistent equations for $m$ and $q$ respectively.
Conversely, by extremizing \ref{freeEn} with respect to $q$, one gets the self-consistent equation for $r$ as
\be \label{erre} 
\langle r (\epsilon,\beta)  \rangle = \lim_{N_H \to \infty} \frac{1}{\epsilon  N_H}\sum_{\mu>1}^{N_B}\frac{q}{(W_{\mu}-\beta(1-q))^2}.
\ee
In the TDL, the last expression can be rewritten through
\begin{equation}\label{eq:Rpw}\small 
\langle r (\epsilon,\beta)  \rangle =  \int \frac{q}{(W^2 - \beta(1-q))^2} \, \frac{1}{W \sqrt{2 \pi} \sigma} e^{- \frac{(\log W - \mu)^2}{2 \sigma^2}}dW,
\end{equation}
where we use eq. \ref{P(W)} and $\mu$ and $\sigma$ are given by eq. \ref{mu} and \ref{sigma}.

A more intuitive route (resembling annealing in spin glasses \cite{MPV}, but ultimately leading to qualitatively correct results), consists in substituting in  eq. \ref{erre} all $W_\mu$ different from $W_1$ ($\mu =1$ is the test-case) with the mean value $\< W \>$.
Explicitly,
\be\nonumber
\langle W \rangle = N_B \langle J \rangle  = N_B \exp[ \langle \chi \rangle_a (\alpha +1) L -L] = N_B^{{\; \gamma\theta- \gamma+1}},
\ee
being $L = \gamma \ln N_B$ \cite{BA1},
in the TDL  three regimes  survive
\be \label{eq:Wann} 
\langle W \rangle =  N_B^{\; \; \gamma\theta- \gamma+1} \to
\begin{cases}
\infty, & \mbox{if} \qquad \theta > 1-\frac{1}{\gamma}, \\
1, & \mbox{if} \qquad \theta = 1-\frac{1}{\gamma} ,\\
0, & \mbox{if} \qquad \theta < 1-\frac{1}{\gamma}.
\end{cases}
\ee
So, when $\langle W \rangle \to \infty$, we can think at the test-clone $B_1$ as \emph{non-self} directed because its connectivity is smaller than the other ones, while when $\langle W \rangle \to 0$ we can think at the test-clone $B_1$ as \emph{self} directed,  being its connectivity  higher than the others.
Accordingly,   $\langle r \rangle$  can assume three different values:
\be \label{eq:R3} 
 \ll r(\epsilon, \beta) \rr =
\begin{cases}
0, & \mbox{if} \qquad \ll W \rr \rightarrow \infty, \\
\frac{q}{(1-\beta(1-q))^2}, & \mbox{if} \qquad \ll W \rr \rightarrow 1, \\
\frac{q}{\beta^2(1-q)^2}, & \mbox{if} \qquad \ll W \rr \rightarrow 0. \\
\end{cases}
\ee
Therefore we can discuss the following three situations:
\begin{enumerate} 
 \item The typical B clone displays $\ll W \rr \to \infty$, namely, it is more connected than the test-clone $B_1$. Thus, $B_1$ can be interpreted as a non-self clone \cite{varela3}. In this case $r$ is vanishing and the self-consistent equation for $m$ is simply
\be\label{eq:Viana}
m^1 = \ll \xi^1 \tanh \big[ \beta \; \frac{m^{1}}{W_1}\xi^{1}\big]\rr_\xi,
\ee
which is the self-consistent equation for an Hopfield model away from saturation  \cite{amit,viana} with a rescaled noise level $\beta'= \beta/W_1$. From an immunological point of view, this means that helpers can successfully exchange signals with the clone $B_1$ under antigenic stimulation.
\item The case $\< W\> \to 1$ has zero probability measure, it recovers the Hopfield neural model near saturation \cite{amit,hopfield}, and can be skipped.
\item The typical B clone displays $\ll W \rr \to 0$. Hence, we can interpret $B_1$ as a self-addressed. The self consistent equations in this case are
\begin{align} 
\label{eq:ANNE}
m^1 &= \langle\langle \xi^1 \tanh \Big[  \beta\Big(\frac{m^{1}\xi^{1}}{W_1}+\frac{\sqrt{\epsilon \, q} \, z}{\beta(1-q)}\Big)\Big]\rangle_\xi \rangle_z , \\ \notag
q &=  \langle\langle \tanh^2 \Big[ \beta\Big(\frac{m^{1}\xi^{1}}{W_1}+\frac{\sqrt{\epsilon \, q} \, z}{\beta(1-q)}\Big)\Big]\rangle_\xi \rangle_z,
\end{align}
where we substituted the expression for $r$ (third equation in \ref{eq:R3}) directly into the equation for $m$ and $q$.

As a result, $B_1$, being much connected, can not feel helper signaling and therefore remain anergic.
\end{enumerate}

\subsection{Breaking of the network performances: the ergodic and the random field thresholds}
To inspect where ergodicity is restored we can start trough the order parameter equation system
$$
q=\langle \tanh^2 \Big[ \beta \Big(  \frac{m_\nu}{W_\nu} + \sqrt{\epsilon r}z \Big)\Big]\rangle_{z,W}
$$
$$
r=\langle \frac{q}{[W-\b(1-q)]^2} \rangle_W
$$
and expand them requiring that $m_{\nu}=0$ at criticality, while the overlap (being a continuous function undergoing a second order phase transition) is small, e.g. $ q=<\b^2 \epsilon r z^2>_z=\epsilon \b^2 q \langle 1/ [W-\b(1-q)]^2 \rangle_W $, then, approximating as usual $\langle f(W) \rangle_W \sim f (\langle W \rangle)$ (annealing) we get the leading term as
$$
q=\frac{\epsilon \b^2}{(<w>-\b)^2}q,
$$
hence
$$
\langle W \rangle = \b(1 + \sqrt{\epsilon}),
$$
which recovers the critical line of the Hopfield model for $\langle W \rangle=1$ as it should.

\subsection{Appendix Three. The mean field biased repertoire: Entropic considerations}

A recently, pioneering experiment, and its analysis trough maximum entropy principle, has revealed a highly non-uniform usage in genes coding for antibodies in zebrafish \cite{zebrafish,bialek}: In particular it has been proven that the sequence distribution follows a Zipf law and there is a massive reduction of diversity, so to say, the repertoire is far from being completely expressed.
As we are going to use a mean-field approximation of this key result, in this section, through standard information theory techniques, we highlight the intimate connection between the size of the antibody's repertoire, its entropy  and the occurring frequency of a single entry.
Recalling that each antibody $\Psi_{\mu}$ is represented as a binary string of length $L$ ($\Psi^i\in(-1,1)$, $i=1\dots L$) whose entries are independent and identically distributed following $P(\Psi)=\prod_{i=1}^LP(\Psi^i)$. Each probability distribution of a dichotomic random variable can be written following  Eq. (\ref{eq:P_rep}),
\begin{equation} \label{eq:P_rep}
P_{\mathrm{rep}}(\Psi_i^{\mu} | a)= \delta(\Psi_i^{\mu}-1) \frac{1+a}{2} + \delta(\Psi_i^{\mu}-1) \frac{1-a}{2},
\end{equation}
where $\delta(x)$ is the Kronecker delta and $a \in [-1,+1]$ tunes the extent of bias, namely the likelihood of repetitions among bitstrings, i.e. $a= \langle \Psi^i \rangle \in (-1,1)$. If we consider the set $A_k=\{\Psi: \sum_{i=1}^L \d_{\Psi^i,1}=k\}$, it is easy to see that
\be
P(A_k)=\binom{L}{k} p^k (1-p)^{L-k}\sim 2^{-L\{S(p)- S(\frac k L)+(\frac k L -p)\log(\frac p {1-p})\}},
\ee
where $p=P(\Psi=1)=(1+a)/2$ and $S(p)$ is the entropy of the probability distribution, defined as
\be
S(p)=-p\log p -(1-p)\log (1-p).
\ee
In the limit $L >> 1$, $P(A_k)$ is non zero only if $k\sim p L$, thus, $A_{pL}$ is  the set of typical strings (having full probability to be drawn). Each typical string $\Psi_{\textrm{typ}}$ has the same probability to occur
\be
P(\Psi_{\textrm{typ}})\sim p^{pL}(1-p)^{(1-p)L}=2^{-L S(p)},
\ee
and the number of typical strings, i.e. the size of the repertoire, is $2^{LS(p)}$. 	When $a=0$ the entropy is maximal and the size of the repertoire is the maximum ($S(1/2)=1$ and $B=2^L$). On the contrary, if $a \neq 0$, the entropy is less than $1$ and the size of the repertoire sensibly decreases.
In a more realistic scenario in which the entries are  not identical distributed \cite{bialek}, we would have different bias parameters $a_i$ for each entry, but the result would be quite the same: as soon as $a_i$ are different from $0$, the size of the repertoire is $2^{L S(a)} << 2^{LS(0)}=2^L$, where this time
\be
S(a)=\frac 1 L\sum_{i=1}^LS(a_i).
\ee
Since we are interested just in reproducing the size of the repertoire, we used the simpler  mean field approximation of the latter, where an effective bias parameter $a$ replaces the whole vector $(a_i)_{i=1}^L$.

\subsection{Appendix Four. Mimicking selection during the ontogenesis of B cells.}
In this section we deepen the simulations performed to mimic the ontogenesis of B cells and the related results.

First, we recall that we model the ensemble of self-molecules as a set $\mathcal{S}$ of strings $\Phi_{\mu}$, of length $L$, whose entries are extracted independently via a proper distribution. The overall number of self-molecules is $|\mathcal{S}| = N_S$, that is, $\mu = 1, ..., N_S$.

We generate $\mathcal{S}$ extracting each string entry $i$ according to the simplest biased-distribution
\begin{equation} \label{eq:P_self}
P_{\mathrm{self}}(\Phi_i^{\mu} | \bar{a})= \delta(\Phi_i^{\mu}-1) \frac{1+\bar{a}}{2} + \delta(\Phi_i^{\mu}-1) \frac{1-\bar{a}}{2},
\end{equation}
where $\delta(x)$ is the Kronocker delta and $\bar{a} \in [-1,1]$.

Then, we generate newborn B cells, represented by the arbitrary string $\Psi$ and accept them whenever Eq.~\ref{eq:vincoli} is fulfilled. We find that within a wide region of the parameters $\chi_P$ and $\chi_N$ the resulting final repertoire $\mathcal{B}$ exhibits a bias. In order to deepen this point we tackle the problem from an analytical perspective trying to corroborate the numerical finding.

We make the following ansatz for the distribution of string entries
\begin{equation} \label{eq:P_rep}
P_{\mathrm{rep}}(\Psi_i^{\mu} | a)= \delta(\Psi_i^{\mu}-1) \frac{1+a}{2} + \delta(\Psi_i^{\mu}-1) \frac{1-a}{2},
\end{equation}
where $a$ can in principle range in $[-1,1]$ and we try to figure out the possible values of $a$ so that all strings extracted via (\ref{eq:P_rep}) fulfill (with probability close to $1$) the constraint in Eq.~\ref{eq:vincoli}. In particular, we aim to figure out any correlation between the parameter $\bar{a}$ (assumed as fixed) and the free parameter $a$.
Notice that the choice of Eq.~\ref{eq:P_rep} is consistent with the results presented in Appendix Three and with our mean-field approach as it provides the easiest distribution, possibly admitting a degree of bias ($a \neq 0$), through which entries are identically and independently generated.

Now, given the REM-like distribution \cite{REM} of complementarities (\ref{eq:compl},\ref{coupling}), in order to estimate $\max_{ s \in \Phi} \{ \chi(\Psi, \Phi)\}$, as suggested in \cite{kosmir1,kosmir2}, one can approximate the extreme value distribution for $\chi(\Psi, \Phi)$ with a Gumbel distribution, whose peak, for large $N_S$, is located at $\langle \chi \rangle_{a,\bar{a}} + \sqrt{2 (\langle \chi^2 \rangle_{a,\bar{a}} - \langle \chi \rangle_{a,\bar{a}}^2) \log N_S}$, where $\langle \cdot \rangle_{a,\bar{a}}$ denotes the average performed over the distributions $P_{\mathrm{self}}(\Phi_i^{\mu} | \bar{a})$ and $P_{\mathrm{rep}}(\Psi_i^{\mu} | a)$, respectively.
%
Recalling Eqs.~\ref{eq:compl}, \ref{eq:P_self}, \ref{eq:P_rep}, we have
\begin{eqnarray}
\langle \chi \rangle_{a,\bar{a}} &=& \frac{L}{2} (1-a \bar{a}) ,\\
\langle \chi^2 \rangle_{a,\bar{a}} - \langle \chi \rangle_{a,\bar{a}}^2 &=&  \frac{L}{2} (1-a^2)(1-\bar{a}^2),
\end{eqnarray}
moreover, as to $N_S$, we can assume the rather general scaling $N_S \approx (N_B)^c$, with $c>0$, largely consistent with
immunogenetics measurements \cite{rob}; thus, we get
\begin{equation} \label{eq:gap}
2 f <  (1-a\bar{a}) + \sqrt{ 2 c \gamma(1-a^2)(1-\bar{a}^2) } < 2 f+ \Delta,
\end{equation}
where $f=\chi_P/L$ and  $\Delta = (\chi_N - \chi_P)2/L$ is the accessible gap (it provides a logarithmic measure of the corresponding allowed binding energies).

In order to fix the value of the parameters, one can rely on indirect measurements, such as the survival probability of new-born B cells, which is expected to be $P_{\textrm{surv}} = 0.1$ (see Fig.~$4$, left panel).
Moreover, we expect that $\chi_P>L/2$, since two randomly generated strings display, on average, $\chi =1/2$, and that $c \gamma$ is relatively small, since the self-repertoire is expected to be sensitively smaller than the B-repertoire \cite{deboer,franz,franz!!!,perelson}.

Having set the parameters according to such constraints, we tune $\bar{a}$ and we accordingly derive the values of $a$ which fulfill the inequality (\ref{eq:gap}), these values are those compatible with the final repertoire.
Interestingly, we find that $a$ and $\bar{a}$ are correlated: positive values of $\bar{a}$ yield a biased mature repertoire with $a>0$.

\subsection{Appendix Five. The robustness of the log-normal connectivity distribution for the idiotypic network}

Each string $\Psi$ has length $L$ and displays, on average, a number $\rho$ of non-null entries distributed according to the binomial $B(\rho | a, L) = \binom{L}{\rho} [(1+a)/2]^{\rho} [(1-a)/2]^{L-\rho}$. In the TDL $N_B \rightarrow \infty$, the string length is divergent and we can approximate the previous distribution with a delta function peaked at the average value $\langle \rho \rangle_a = (1+a)L/2$.

The observable $\chi_{\mu \nu}$  represents the number of complementarities between two generic strings $\Psi_{\mu}, \Psi_{\nu} \in \mathcal{B}$, defined as
\be
\chi_{\mu \nu}= \sum_{k}^L [\Psi_{\mu}^k(1-\Psi_{\nu}^k) + \Psi_{\nu}^k(1-\Psi_{\mu}^k) ],
\ee
and has the expected values
\begin{eqnarray}
\label{eq:barchi}
\langle \chi \rangle_a &=& \frac{1-a^2}{2}\\
\langle \chi^2 \rangle_a &=& \frac{(1-a^2)^2}{4} \frac{L}{L-1} = \langle \chi \rangle^2  \frac{L}{L-1}
\end{eqnarray}
over the distribution $B(\rho | a, L)$. Notice that, in the TDL, the variance is vanishing and this distribution also converges to a delta peaked at $\langle \chi \rangle_a$.
Hence, exploiting CLT, the stochastic variable $\chi$ can be thought of as normally distributed with $\mathcal{N}(\langle \chi \rangle_a, \langle \chi \rangle_a^2/L))$.

From $\chi_{\mu \nu}$ we can define more precisely the coupling strength $J_{\mu \nu}$ as
\be
J_{\mu \nu} = e^{\alpha \chi_{\mu\nu} - (L - \chi_{\mu\nu})},
\ee
where positive (complementary matches) and negative (non-complementary matches) contributions to the coupling have been highlighted.
The term $\exp(\chi)$ is, by definition, distributed according to the log-normal distribution $\log \mathcal{N}(\langle \chi \rangle_a , \langle \chi \rangle_a^2/L)$. With slight algebraic manipulations, we get that $J$ is distributed according to $\log \mathcal{N}(\langle \chi \rangle_a(\alpha+1)L - L , \langle \chi \rangle_a^2 (\alpha+1)^2 L)$, whose probability distribution is
\begin{equation}
P_{L,N_B}(J | a,L,\alpha) = \frac{1}{J \sqrt{2 \pi L} \langle \chi \rangle_a (\alpha +1)} e^{- \frac{[\log J - \langle \chi \rangle_a(\alpha+1)L + L]^2}{2 \langle \chi \rangle_a^2 (\alpha+1)^2 L }}.
\end{equation}
%
Recalling that $L= \gamma \log N_B$, we can write
\begin{eqnarray}
\langle J \rangle_a &=& N_B^{[\theta^2 + 2 \theta -2]\gamma/2},\\
\langle J^2 \rangle_a &=& N_B^{2 [\theta^2 + \theta -1]\gamma},
\end{eqnarray}
being $\theta = \langle \chi \rangle_a (\alpha + 1) \geq 0$.

We notice that, by properly tuning $a$ and $\alpha$, one can recover, in the TDL, different regimes characterized by different behaviors (finite, vanishing or diverging) for the average $E(J) \equiv \langle J \rangle_a$ and the variance $V(J) \equiv \langle J^2 \rangle_a -\langle J^2 \rangle_a^2$, respectively (see Fig.~$5$).

It is worth stressing that a vanishing $\langle  J \rangle_a$ does not necessarily imply that the emerging topology is under-percolated. This remains true even assuming the stronger condition \cite{BA1} $J _{\mu \nu} = \Theta[\chi_{\mu \nu}(\alpha+1)-1] \exp[\chi_{\mu \nu}(\alpha+1)L-L]$, being $\Theta$ the Heaviside function.

Let us now consider the weighted degree $W$ and its distribution $P(W|a,\alpha,N_B)$. First, we notice that $W$ is a sum of log-normal variables, pairwise not correlated (as their corresponding receptors are independently extracted through random VDJ reshuffling \cite{bialek}).
Then, $W$ can be well approximated by a new log-normal random variable $\hat{W} = \exp({\hat{\chi}})$, where $\hat{\chi}$ is a Gaussian random variable with mean $\mu$ and variance $\sigma^2$. As a result, we expect $\langle \hat{W} \rangle_a = \exp(\mu + \sigma^2/2)$ and $\langle \hat{W}^2 \rangle_a= \exp(2\mu + \sigma^2)$.
Moreover, we can write $\langle W \rangle_a \approx N_B \langle J \rangle_a$ and $\langle W^2 \rangle_a -\langle W \rangle_a^2 \approx N_B (\langle J^2 \rangle_a - \langle J \rangle_a^2)$, in agreement with Bienayme's theorem. Now we can use the previous expressions to fix $\mu$ and $\sigma^2$, recovering the Fenton-Wilkinson
method for approximating log-normal sums.
%
where $E(J) \equiv \langle J\rangle_a = N_B^{[\theta^2 + 2 \theta -2]\gamma/2}$, $\langle J^2 \rangle_a = N_B^{2 [\theta^2 + \theta -1]\gamma}$, being $\theta = (1 - a^2)/2 (\alpha + 1) L \geq 0$. Consequently, by properly tuning $a$ and $\alpha$, one can recover, in the thermodynamic limit, different regimes characterized by different behaviors (finite, vanishing or diverging) for the average $\langle J \rangle_a$ and the variance $V(J) \equiv \langle J^2\rangle_a - \langle J \rangle_a^2$, respectively, as reported in Fig.~$5$.
\newline
In particular, we can write
\begin{eqnarray} \label{mu}
\mu &=& \log \left[ \frac{ N_B \langle J \rangle_a^2 }{ \sqrt{ \langle J \rangle_a^2 +  (\langle J^2 \rangle_a - \langle J \rangle_a^2)/N_B}}  \right],\\ \label{sigma}
\sigma^2 &=&\log \left[ \frac{  \langle J^2 \rangle_a - N_B \langle J \rangle_a^2 }{N_B \langle J \rangle_a^2 } +2 \right],
\end{eqnarray}
through which we get the following distribution for $\hat{W}$, to be taken also as an approximation for $P(W| a, \alpha, N_B)$
\begin{equation}\label{P(W)}
P(W) = \frac{1}{W \sqrt{2 \pi} \sigma} e^{- \frac{(\log W - \mu)^2}{2 \sigma^2}}.
\end{equation}
These results are corroborated by numerical data.
\newline
Therefore, $W$ is characterized by mean and variance which may assume a vanishing, or finite, or diverging value according to the value of $a$ and $\alpha$, similarly to what found for $J$.

\section*{Acknowledgments}
This research was sponsored by the FIRB grant RBFR08EKEV.\\
Sapienza Universit\`{a} di Roma and INFN are acknowledged too.


\end{document}